\documentclass[pre,amsmath,twocolumn,a4paper,showpacs]{revtex4}

\usepackage{amssymb}
\usepackage{graphicx}
\usepackage{natbib}
\usepackage{color}

\newcommand{\mT}{{\mathcal{T}}}
\newcommand{\Dt}{{D_t}}
\newcommand{\ddt}{{{\hat D}_t}}
\newcommand{\ee}{\end{equation}}
\newcommand{\be}{\begin{equation}}
\newcommand{\eea}{\end{eqnarray}}
\newcommand{\bea}{\begin{eqnarray}}
\newcommand{\ea}{\end{array}}
\newcommand{\chg}[1]{#1}

\begin{document}


\title{Stochastic thermodynamics of a tagged particle within a harmonic chain}

\author{David Lacoste$^1$}
\author{Michael A. Lomholt$^2$}

\affiliation{$^1$ Laboratoire de Physico-Chimie Th\'eorique - UMR CNRS Gulliver 7083,\\ PSL Research University, ESPCI, 10 rue Vauquelin, F-75231 Paris, France\\
$^2$ MEMPHYS - Center for Biomembrane Physics, Department of Physics, Chemistry and Pharmacy, University of Southern Denmark, Campusvej 55, 5230 Odense M, Denmark}

\date{\today}

\begin{abstract}
We study the stochastic thermodynamics of an overdamped harmonic chain, which can 
be viewed equivalently as a 1D Rouse chain or as an approximate model of single file diffusion. 
We discuss mainly two levels of description of this system: the Markovian
level for which the trajectories of all the particles of the chain are known and the non-Markovian level
in which only the motion of a tagged particle is available. For each case, 
we analyze the energy dissipation and its dependence on initial conditions. Surprisingly, we find that the average  
 coarse-grained entropy production rate can become \chg{transiently} negative 
when an oscillating force is applied to the tagged particle.
 This occurs due to memory effects as shown in a framework based on path integrals or on a generalized
 Langevin equation.
\end{abstract}


\maketitle


\section{Introduction}

The study of the fluctuations of a tracer particle in a complex medium,
often referred to as single particle tracking is becoming a central tool 
in the study of biological systems. Depending on the environment and the tracer particle, 
standard diffusion, anomalous or super-diffusion can be observed in such systems.

In order to understand quantitatively this diversity of possible behaviors,
model systems play a central role. One of such models is the one
dimensional overdamped motion of non-passing particles (so called
single file diffusion), in which a tracer particle behaves 
sub-diffusively. About 50 years of theoretical work have been devoted to this
rich problem by various methods \cite{Harris1965,Percus1974,Alexander1978,Beijeren1983,Kollmann2003},
and this effort is continuing even today. Among many recent studies, let us 
mention in particular derivations of the large deviation function of the position of a tagged particle 
in single file diffusion \chg{\cite{Krapivsky2014},\cite{Hegde2014a}, and a 
study of the statistics of the tagged particle position, when it is 
subject to a bias in a dense diffusive single file environment \cite{Illien2013}}.

In order to simplify the complex many body features of single file diffusion, 
various approximations have been introduced. One of them is the harmonization method, 
which amounts to replacing the non-linear interactions between neighboring particles 
by harmonic interactions with a spring constant calculated from the thermodynamical properties of the particle system \cite{lizana10}.
This harmonization renders the model easily solvable while still capturing 
the long time and long wavelength properties of the original many body problem,
such as the sub-diffusive behavior and the lasting effect of the initial condition for the tagged 
particle dynamics. The latter
effect, which results from the long memory present in this system, has been studied more systematically in \cite{leibovich13} 
for the original single file problem.
Such memory effects arise in the effective non-Markovian dynamics and result from integrating out all the degrees of 
freedom of the chain except that of the 
tagged particle. These effects have also been studied recently 
in the finite Rouse chain both at the underdamped and overdamped level \cite{Maes2013a}. 

In a separate direction, much progress has been accomplished recently in the statistical physics
of non-equilibrium systems. Stochastic thermodynamics has emerged as a new branch
of thermodynamics specially aimed at describing the time evolution of thermodynamic quantities, like
work, heat and entropy \cite{Ritort2008_vol137,Jarzynski2011_vol2,Seifert2012}. 
In small systems, the fluctuations of these quantities can be measured and characterized. 
The corresponding probability distributions 
satisfy general symmetry relations known
as fluctuation theorems \cite{Crooks2000_vol61,Hatano2001_vol86}, 
which are symmetry properties of large deviation functions. 
Most applications of stochastic thermodynamics have considered 
systems with a small number of degrees of freedom, so that specific studies of non-equilibrium 
systems containing a large number of degrees of freedom with this framework 
still remains largely unexplored.  

In this context, the present paper represents an attempt to use the framework of stochastic thermodynamics
for a simple model of this kind, namely an overdamped harmonic chain.  This paper is composed of the following sections:
in section \ref{sec:II}, we introduce a continuous description of an overdamped harmonic chain, and we 
discuss thermodynamic quantities at this level; in section \ref{sec:III}, we introduce a variant of this model with a finite  
number of particles; in section \ref{sec:IV}, we study a non-Markovian level of description, 
constructed by keeping only the position of a tagged particle as dynamic variable and integrating over the 
other degrees of freedom, while in section \ref{sec:V}, we study a related case, 
constructed by keeping only the positions of two tagged particles in the chain.

\section{The Markovian description}\label{sec:II}

The physical system we study is that of hard-core interacting overdamped particles distributed uniformly in one dimension. 
We mainly work in a continuum limit where the particle positions are described by a field $x(n,t)$ with $t$ being time and $n$ 
being the particle number along the infinite chain, i.e., it ranges from $n=-\infty$ to $n=\infty$. It was shown in \cite{lizana10} that this system is mathematically equivalent to the Edward-Wilkinson chain. Thus our starting point is the equation of motion
\begin{equation}
\xi \frac{\partial x(n,t) }{ \partial t} = \kappa \frac{ \partial^2 x(n,t) }{ \partial n^2} + f_\mT(t)\delta(n) +  \eta(n,t).
\label{Langevin} 
\end{equation}
Here $\xi$ is the friction constant of each particle, which is related to their diffusion constant $D$ by $\xi=k_B T/D$ with $k_B T$ being Boltzmann's constant times temperature. $\kappa$ is an effective spring constant between nearest neighbor particles in the system. It is related to the isothermal compressibility $\chi_T$ of the system by $\kappa=\rho/\chi_T$, where $\rho$ is the average density of particles. For a system of hard-core interacting point particles it is given by $k_B T \rho^2$. The force $f_\mT(t)$ is an external time-dependent force that acts on particle number $n=0$, which we will also call the tagged particle denoting it $x_\mT(t)=x(n=0,t)$. Finally, the noise $\eta(n,t)$ arising from the surrounding heat bath is Gaussian with zero mean and correlations
\begin{equation}
\langle \eta(n,t)\eta(n',t')\rangle = 2 k_B T \xi \delta(t-t')\delta(n-n').
\end{equation}
For Eq. (\ref{Langevin}) to be a valid description of single file diffusion we assume that we study time scales that are long enough for neighbouring particles to have interacted, i.e., that $t\gg 1/(\rho^2 D)$.

Since we are interested in studying this system out of equilibrium, 
we will take the chain to be initially at equilibrium with a heat bath of temperature $T_{\rm chain}$, 
which can be different from the temperature $T$ that the surrounding heat bath has at $t>0$. 

One consequence of this is that the average initial positions of the particles are $\langle x(n,t=0)\rangle=\langle x_\mT(0)\rangle+n/\rho$. 
To avoid this 
non-zero average we will find it convenient in the following to work with the position relative to this average:
\begin{equation}
y(n,t)=x(n,t)-\langle x_\mT(0)\rangle-n/\rho.
\end{equation}
Note that in the $y$ coordinates our system is equivalent to the Rouse chain in one dimension.

To work out the consequences of our non-equilibrium initial condition on the motion of the tagged particle we can Fourier and Laplace transform Eq. (\ref{Langevin}). Denoting the transform by a change of variable, e.g. $y(q,s)=\int_{-\infty}^\infty d n\int_0^\infty d t\,e^{-i q n-s t} y(n,t)$, we can isolate $y$ and find
\begin{equation}
y(q,s)=\frac{\eta(q,s)+\xi y(q,t=0)+f_\mT(s)}{\xi s+\kappa q^2}.
\label{eqofmotion} 
\end{equation}
Taking $f_\mT=0$ for now we will have that the average position of the tagged particle remains at the origin. Following Appendix A of \cite{lizana10} we will write the covariance as
\begin{equation}
\langle y(q,s)y(q',s')\rangle = A_{\rm init}(q,q',s,s')+A_{\rm noise}(q,q',s,s')
\end{equation}
where
\begin{align}
&A_{\rm init}(q,q',s,s')=\frac{\langle y(q,t=0) y(q',t=0)\rangle}  {(s+\kappa q^2/\xi)(s'+\kappa q'^2/\xi)},\label{eq:Ainit}\\
&A_{\rm noise}(q,q',s,s')= \frac{ \langle \eta(q,s) \eta(q',s') \rangle}  {(\xi s+\kappa q^2)(\xi s'+\kappa q'^2)},
\end{align}
To calculate $A_{\rm noise}$ we note that the Fourier and Laplace transform of the correlation of the noise is
\begin{equation}
\langle \eta(q,s) \eta(q',s') \rangle = \frac{4\pi\xi k_BT \delta(q+q')}{s+s'}.
\end{equation}
Therefore
\begin{equation}
A_{\rm noise}(q,q',s,s')=  
\frac{4\pi\xi k_BT \delta(q+q')}
{(s+s')(\xi s+\kappa q^2)(\xi s'+\kappa q^2)}.
\end{equation}
and inverting the Laplace transforms we have
\begin{equation}
A_{\rm noise}(q,q',t,t') =
2\pi \delta(q+q')
\frac{e^{-\kappa q^2|t-t'|/\xi} - e^{-\kappa q^2(t+t')/\xi} }{\kappa q^2/(k_B T)}.\label{eq:Anoiseres}
\end{equation}
From this expression we can obtain $\langle y(q,t=0) y(q',t=0)\rangle$ by noting that the infinite time limit of $A_{\rm noise}(q,q',t,t)$ gives the equilibrium covariance $\langle y(q)y(q')\rangle$ at temperature $T$, except for the point $q=q'=0$ (which is determined by the condition of the tagged particle beginning at the origin - see later). Replacing $T$ with the actual temperature $T_{\rm chain}$ of the chain at $t=0$ we then find (for $(q,q')\ne (0,0)$)
\begin{align}
\langle y(q,t=0) y(q',t=0)\rangle&=\lim_{t\to \infty}\left.A_{\rm noise}(q,q',t,t)\right|_{T=T_{\rm chain}}\nonumber\\
&=2\pi\delta(q+q')\frac{k_B T_{\rm chain}}{\kappa q^2}
\label{equipart}
\end{align}
After inverting the Laplace transforms in Eq. (\ref{eq:Ainit}) we can therefore write the correct result including the point $(q,q')=(0,0)$ formally as
\begin{align}
A_{\rm init}(q,q',t,t') =&
e^{-\kappa q^2 (t+t')/\xi}\, 2\pi \delta(q+q')
\frac{k_B T_{\rm chain}}{\kappa q^2}\nonumber\\
&-(2\pi)^2\delta(q)\delta(q')\int\frac{d q''}{2\pi}\frac{k_B T_{\rm chain}}{\kappa {q''}^2}
\label{eq:Ainitres}
\end{align}
where the last term comes from the condition that $A_{\rm init}(n=0,n=0,t=0,t=0)=0$ such that the tagged particle is at the origin at $t=0$.
In expressions such as Eq. (\ref{eq:Ainitres}) where the infinite continuum description results in diverging integrals over $q$ 
we have to regularize the expression by cutoffs, i.e., a lower limit on the $q$-integral at $q_{\rm min}\sim 2\pi/L$ where $L$ is the length of the system and an upper limit $q_{\rm max}\sim 2\pi\rho$.

Having worked out the covariances of $y$ we can now obtain for instance the mean square displacement (MSD) $\langle \delta x_\mT^2(t)\rangle=\langle [x_\mT(t)-x_\mT(0)]^2\rangle$ of the tagged particle by inverting the Fourier transforms at $n=n'=0$. 
The result in the present case without external force is
\begin{align}
\langle \delta x_\mT^2(t)\rangle_{f=0} &= \langle y(n=0,t)^2\rangle_{f=0} \nonumber\\
&=k_B\left(T+T_{\rm chain}(\sqrt{2}-1)\right)\sqrt{\frac{2 t}{\pi \xi \kappa}}\label{eq:effMSD}
\end{align}
Notice that the result agrees with the $1/\sqrt{2}$ slowdown obtained in \cite{lizana10,leibovich13} for the case of 
initially equidistant particles corresponding to $T_{\rm chain}=0$ compared with the MSD in thermal equilibrium where $T_{\rm chain}=T$. Introducing the temperature $T_{\rm chain}$ allows us to treat the two cases of equidistant particles or thermal equilibrium simultaneously as one case. 
However, notice that if the original system (before harmonization) consisted of particles with only hard-core repulsion, then the equilibrium distribution of the original system is independent of temperature, since for that interaction the particles will have the same Poissonian distribution in equilibrium no matter the temperature.

\subsection{The first law of thermodynamics}
Since the equation of motion of the chain is the overdamped Langevin equation without inertia, Eq. (\ref{Langevin}), 
the internal energy of the chain is equal to its potential energy, namely: 
\begin{align}
U(t) &= \frac{\kappa}{2}\int d n\, \left(\frac{\partial x}{\partial n}-\rho^{-1}\right)^2, \nonumber\\
&=\frac{\kappa}{2}\int \frac{d q}{2\pi}\, q^2 y(q,t)y(-q,t)
\label{internal energy}
\end{align}
The system is driven out of equilibrium by a time-dependent force, which
is applied only on the tagged particle at the origin:  
$f(n,t)= \delta(n) f_\mT(t)$.
As a result, the stochastic work is 
\begin{align}
W &=\int_0^t d t'\, \int dn\, f(n,t') \dot{x}(n,t'), \nonumber\\
  &=\int_0^t d t'\, f_\mT(t') \dot{x}_\mT(t'), 
 \label{work} 
\end{align}
while the heat is \cite{Sekimoto1998}
\begin{align}
Q  &= \int_0^t d t'\, \int dn [ -\eta(n,t')+\xi \dot{x}(n,t') ] \dot{x}(n,t'), \nonumber \\
   &=\int_0^t d t'\, \int dn [ \kappa \frac{\partial^2 x(n,t')}{\partial n^2} +\delta(n) f_\mT(t') ] \dot{x}(n,t'). 
\label{heat}
\end{align}
Note that we have used the Langevin equation, Eq. (\ref{Langevin}), to go from the first to second line.
It is also important to point out that the stochastic integral must be interpreted with the Stratonovich convention.

With these definitions, the first law at the trajectory level 
has the form $W-Q=\Delta U$, where $\Delta U=U(t)-U(0)$.
The sign convention is such that work is counted as positive when work is done
on the particle and heat is counted as positive when given to the environment. 
 
\subsection{Averages of thermodynamic quantities}
Averaging over noise and inverting in Laplace space, we find from Eq. (\ref{eqofmotion})
\begin{equation}
\langle y(q,t)\rangle =\xi^{-1}\int_0^t d t'\, e^{-\kappa q^2(t-t')/\xi} f_\mT(t')
\end{equation}
We can Fourier invert this to find
\begin{equation}
\langle y(n,t)\rangle =\xi^{-1}\int_0^t d t' \frac{e^{-n^2/(4\kappa(t-t')/\xi)}}{\sqrt{4\pi\kappa(t-t')/\xi}}f_\mT(t').
\label{x(n,t)}
\end{equation}
From Eq. (\ref{work}), one then obtains the average work on the chain as
\begin{align}
\langle W\rangle&=\int_0^t d t'\, f_\mT(t')\langle \dot{x}_\mT(t')\rangle\nonumber\\
&=\int_0^t d t'\, f_\mT(t')\frac{d}{d t'}\int_0^{t'}d t''\frac{f_\mT(t'')}{\sqrt{4\pi\kappa\xi(t'-t'')}}
\label{average work} 
\end{align}

When we take the ensemble average of the internal energy $U$, we can divide the result into 
two contributions: one arises purely due to the thermal noise in the system, $U_{f=0}$, and the other is the additional energy due to 
the external force, $\delta U_{f}$. 
For the external force contribution, we find from the term linear in force in Eq. (\ref{eqofmotion}) that it has average
\begin{equation}
\langle \delta U_f(t) \rangle = \frac{1}{8\sqrt{\kappa\xi\pi}}\int_0^t d t'\int_0^{t}d t''\frac{f_\mT(t')f_\mT(t'')}{(2t-t'-t'')^{3/2}}.
\end{equation}

Let us consider now the particular case of a constant force turned on at $t=0$, i.e., $f_\mT(t)=F \Theta(t)$, where $\Theta$ is the Heaviside function. 
Eq. (\ref{work})
immediately tells us that then 
\be
W=F (x_\mT(t)-x_\mT(0) ),
\label{work-cte-force}
\ee
From Eq. (\ref{x(n,t)}) or Eq. (\ref{average work}), one obtains  
\begin{equation}
\langle W \rangle = \frac{F^2 \sqrt{t}}{\sqrt{\pi \kappa \xi}}.
\label{average-work}
\end{equation}
Note that a consequence of this peculiar scaling of the work with the time $t$ is that 
the average work per unit time vanishes in the limit $t \rightarrow \infty$.

Now, the contribution due to the external force in the average internal energy is
\begin{align}
\langle \delta U_f (t)\rangle & = \frac{F^2 \sqrt{t}}{8 \sqrt{\xi \kappa \pi}} \left( 8 - 4 \sqrt{2} \right), \nonumber\\
&= \langle W\rangle \left( 1 -  \frac{\sqrt{2} }{2} \right),
\end{align}
For the contribution due to thermal fluctuations one should instead consider the energy density (i.e. energy per particle) 
$u(n,t)=\kappa (\partial y/ \partial n)^2/2$, rather than the global energy summed over all particles, namely $U_{f=0}(t)$ 
which is divergent in this infinite system. 
From our expressions for Eqs. (\ref{eq:Anoiseres}) and (\ref{eq:Ainitres}) we find
\begin{equation}
\langle u_{f=0}(n,t) \rangle = \frac{k_B T}{2} \int \frac{dq}{2 \pi} \left[ 1 - \left(1-\frac{T_{\rm chain}}{T}\right)e^{-2 \kappa q^2 t /\xi} \right],
\end{equation}
This result diverges due to the infinitely many degrees of freedom in the continuum description. 
We can cure the singularity at $t=\infty$ by subtracting the energy density at that point in time to find
\begin{align}
\langle u_{f=0}(n,t) -u_{f=0}(n,t=\infty) \rangle &= \frac{k_B (T_{\rm chain}-T)}{4 \sqrt{2 \pi \kappa t/\xi}}.\label{eq:uaverage} 
\end{align}
We will check in Section \ref{sec:IV} that this result also emerges from taking the appropriate limit for the internal energy of a finite discrete chain.

\subsection{Probability distribution and large deviation of the work}
Beyond averages, the thermodynamic quantities introduced above have well-defined distributions: 
the distribution of work has been first studied in the context of Rouse polymers in Ref.~\cite{Speck2005}
at a Markovian level. This quantity was then studied in the context of non-Markovian processes, with long-time memory,
in Ref.~\cite{Chechkin2009}. Here, 
since the work is a linear function of $x(n,t)$ which is a Gaussian variable, it is itself a Gaussian variable. 
In the constant force case, the average work is given in Eq. (\ref{average-work}), while the variance of the work, $\sigma_W^2$,
follows from Eqs. (\ref{eq:effMSD}) and (\ref{work-cte-force}). Indeed,  
\bea
\sigma_W^2 &=& \langle W^2 \rangle - \langle W \rangle^2, \nonumber\\
&=& F^2 \langle \delta x_\mT^2 \rangle_{f=0} \nonumber\\
&=& F^2 k_B \left[T+T_{\rm chain}(\sqrt{2}-1)\right] \sqrt{ \frac{2t}{\pi \kappa \xi}},
\label{variance-work}
\eea
By linearity of the equations of motion, it follows that the variable $x_\mT - \langle x_\mT \rangle$ satisfies the same equations as $x_\mT$ in
the absence of an applied force. This is why we can use $\langle \delta x_\mT^2 \rangle_{f=0}$ in the second line of the above equations.

In this problem, the probability density function of the work $P(W)$ should satisfy the Crooks fluctuation relation provided the initial
condition corresponds to an equilibrium one, i.e., $T_{\rm chain}=T$ \cite{Chechkin2009}. 
In the present case the relation can be written as \cite{Crooks2000_vol61}
\be
\frac{P(W)}{P(-W)}=e^{\beta( W-\Delta \mathcal{F})},
\label{Crooks}
\ee
where $\beta=1/k_BT$ and $\Delta \mathcal{F}$ is the free energy difference evaluated between the initial and final potentials at thermal equilibrium. 
Since the force $F$ is assumed to be external, there is no change in potential energy in such conditions, 
and thus $\Delta \mathcal{F}=0$ in the present case.
In the particular case 
of Gaussian distributions, this fluctuation relation is equivalent to
\be
\langle W \rangle - \Delta \mathcal{F}=\frac{\sigma_W^2}{2 k_B T} ,
\label{FT-work}
\ee

From Eqs. (\ref{average-work}) and (\ref{variance-work}), we see that Eq. (\ref{FT-work}) is indeed satisfied
for any constant force $F$ and any time $t$ provided that the initial condition is at equilibrium, $T=T_{\rm chain}$.
If the initial condition is not at equilibrium
then the Crooks relation does not hold.

It is interesting to note that while the transient fluctuation relation holds at any time $t$, 
the work does not obey a large deviation principle in the usual way, i.e. in a way which is linear in time.
Indeed, although the work is Gaussian, 
\be
P(W) = \frac{1}{\sqrt{2 \pi \sigma_W^2}} e^{-\left( W - \langle W \rangle \right)^2/2 \sigma_W^2},
\ee
its first and second moments scale as $\sqrt{t}$, which means that the large deviation 
of the work should be defined with respect to $\sqrt{t}$ as follows:
\be
P({W}/{\sqrt{t}}=\omega) \simeq e^{- \left( \omega \sqrt{t} -\langle W \rangle \right)^2 / 2 \sigma_W^2} 
\simeq e^{-\alpha \sqrt{t}},
\ee
where $\alpha$ is the constant
\begin{equation}
\alpha=\frac{\sqrt{2\pi \kappa \xi} (\omega -F^2/\sqrt{\pi \kappa \xi})^2}{4 F^2 k_B [T+T_{\rm chain}(\sqrt{2}-1)]}.
\end{equation}
Despite this special scaling of the large deviations, positive and negative large deviations of the work 
are still for $T_{\rm chain}=T$ connected by the same fluctuation relation:
\be
\frac{P(\frac{W}{\sqrt{t}}=\omega)}{P(\frac{W}{\sqrt{t}}=-\omega)}=e^{\beta \omega \sqrt{t}}\quad (T_{\rm chain}=T),
\ee
in agreement with Crooks relation of Eq. (\ref{Crooks}).
Therefore, we see that a uniform large deviation principle in $\exp{(-\alpha t)}$ is not needed 
for the transient fluctuation relations to hold. 
For a more detailed discussion of this point, 
we refer to a recent study of a Maxwell-Lorentz particle model \cite{Gradenigo2013}. 

For the initial condition of equidistant positions, $T_{\rm chain}=0$, the variance $\sigma_W^2$ is reduced
by a factor $\sqrt{2}$ and therefore the constant $\alpha$ is increased by the same amount since 
the other factors such as the average work do not depend on the initial condition.
As also mentioned above, the Crooks relation does not hold when the initial condition is not at equilibrium. 
Instead we have 
\be
\frac{P(\frac{W}{\sqrt{t}}=\omega)}{P(\frac{W}{\sqrt{t}}=-\omega)}=e^{\sqrt{2} \beta \omega \sqrt{t}}\quad (T_{\rm chain}=0).
\ee
Finally, we note that non-Gaussian large deviations for the position of the tagged particle in single file diffusion 
(with the same scaling in time and the same dependence on the initial
condition) have been obtained recently using Macroscopic Fluctuation Theory \cite{Krapivsky2014}  \chg{and in \cite{Hegde2014a}
by a different method}
without relying on the harmonization approximation considered in the present work.

\section{The discrete chain}\label{sec:III}

To gain further insight in the thermodynamics of our system we investigate in this section a finite discrete chain. 
This allows us to calculate the internal energy and the entropy of the chain without having to regularize integrals.
 
In this discrete version of the system we will label the particle positions by $x_n$ and assume that they move inside a ring 
of circumference $L$. Besides harmonization of the interaction between particles then each particle $n$ is assumed to be bound 
to the position $n/\rho$ by a harmonic spring with potential $\tfrac{1}{2}k(x_n-n/\rho)^2$. This spring will be taken to remember 
how many times it has been wound around the ring. This means that each coordinate can range from $-\infty$ to $\infty$ and that it 
will be a Gaussian random variable due to the linearity of our model. Furthermore, our model will have a well defined equilibrium 
with a Gaussian probability distribution of the positions due to the additional harmonic potential on positions. But for quantities 
for which it is convenient we can always eliminate this potential by taking $k\to 0$. Our system of Langevin equations for the particle 
positions are then
\begin{equation}
\xi {\dot x}_n=\kappa\left(x_{n+1}+x_{n-1}-2 x_n\right)-k(x_n-n/\rho)+f_n+\eta_n
\end{equation}
where $f_n$ and $\eta_n$ are the discrete counterparts of external force and thermal noise on particle $n$ with the noise correlation $\langle \eta_n(t)\eta_m(t')\rangle=2 k_B T \xi \delta(t-t') \delta_{n,m}$. We let the index $n$ run over the particles from $-M$ to $M$ with the total number of particles then being $N=2M+1$. Furthermore we use the notation $x_{M+1}=x_{-M}+L$ and $x_{-M-1}=x_M-L$. 

For notational convenience we introduce similarly to the continuum case the coordinates $y_n=x_n-n/\rho$, as well as a matrix $A$ with components $A_{mn}=2 \delta_{m,n}-\delta_{n,m-1}-\delta_{n,m+1}$. Collecting the coordinates $y_n$ in a column vector $y$ we can then write for instance the internal energy as
\begin{align}
U&=\frac{\kappa}{2}\sum_{n=-M}^M(y_n-y_{n-1})^2+\frac{k}{2}\sum_{n=-M}^M y_n^2\\
&=\frac{\kappa}{2}y^T A y+\frac{k}{2}y^T y
\end{align}
where ${}^T$ denotes transposition.

To decouple the degrees of freedom in our system we will do a discrete Fourier transform of the coordinates
\begin{equation}
{\tilde y}_n=\frac{1}{\sqrt{N}}\sum_{m=-M}^M y_m e^{i m u_n}\; ,\quad u_n=2\pi n/N
\end{equation}
where the index $n$ can also be taken to range from $-M$ to $M$. This transform can alternatively be written
\begin{equation}
{\tilde y}=\Lambda^\dagger y
\end{equation}
where the matrix $\Lambda^\dagger$ is the complex conjugated and transposed (and thereby inverse) of the unitary matrix $\Lambda$ with components
\begin{equation}
\Lambda_{mn}=\frac{1}{\sqrt{N}} e^{-i m u_n}
\end{equation}
This transformation decouples the system since the corresponding change of basis leads to a diagonal matrix ${\tilde A}=\Lambda^\dagger A \Lambda$ with components
\begin{equation}
{\tilde A}_{mn}=2(1-\cos u_n) \delta_{m,n}
\end{equation}
and therefore we have a Langevin equation in the tilted basis which is
\begin{equation}
\xi {\dot {\tilde y}}_n=-(\kappa {\tilde A}_{nn}+k){\tilde y}_{n}+{\tilde f}_n+{\tilde \eta}_n
\end{equation}
If we assume again that the particles initially have a distribution which corresponds to equilibrium at a temperature $T_{\rm chain}$ then this tells us that ${\tilde y}_n$ will be Gaussian variables with average
\begin{equation}
{\tilde m}_n(t)\equiv \langle {\tilde y}_n(t)\rangle = \int_0^t d t' e^{-(t-t')/\tau_n} {\tilde f}_n(t')/\xi
\end{equation}
with the time-scale $\tau_n^{-1}=(\kappa {\tilde A}_{nn}+k)/\xi$ and with a covariance matrix which is diagonal with a diagonal that are the variances:
\begin{equation}
{\tilde V}_{nn}(t)=\frac{k_B T}{\xi}\tau_n+\frac{k_B(T_{\rm chain}-T)}{\xi}\tau_n e^{-2 t/\tau_n}
\end{equation}

From the above decoupled solution a number of quantities can be calculated. For instance, the MSD of the tagged particle is
\begin{equation}
\langle \delta x_\mT^2(t)\rangle =\langle [y_0(t)-y_0(0)]^2\rangle
\end{equation}
To calculate this we note that in general
(${}^*$ denotes complex conjugation)
\begin{equation}
\langle y_0(t_2)y_0(t_1)\rangle= \sum_{m,n} \Lambda_{0m} \Lambda_{0n}^* \langle {\tilde y}_m(t_2){\tilde y}_n(t_1)^*\rangle
\end{equation}
and that the two-time correlation matrix for $t_2>t_1$ in the absence of external force is
\begin{equation}
\langle {\tilde y}_m(t_2){\tilde y}_n(t_1)^*\rangle = e^{-(t_2-t_1)/\tau_n}{\tilde V}_{nn}(t_1)\delta_{m,n}
\end{equation}
Inserting this and using $|\Lambda_{mn}|^2=1/N$ we get
\begin{align}
&\langle \delta x_\mT^2(t)\rangle = \frac{1}{N}\sum_n\left[{\tilde V}_{nn}(t)+{\tilde V}_{nn}(0)-2{\tilde V}_{nn}(0)e^{-t/\tau_n}\right]\nonumber\\
&\;=\frac{2 k_B T}{N \xi}\sum_n\tau_n\left[1-e^{-t/\tau_n}+\frac{T_{\rm chain}-T}{2 T}(1-e^{-t/\tau_n})^2\right]\label{eq:exactMSD}
\end{align}
From this formula we can easily extract three limits. The first is for $t\ll \xi/\kappa$ where $t/\tau_n \ll 1$ for all $n$ (we take $k\to 0$ here). Expanding to first order in $t/\tau_n$ we get
\begin{equation}
\langle \delta x_\mT^2(t)\rangle \sim 2 D t\; ,\quad t\ll \xi/\kappa
\end{equation}
i.e., the tagged particle behaves like a free particle. In the other extreme, where we take $t\gg \tau_n$ for all $n$, i.e., $t\gg N^2 \xi/\kappa$, 
we get that all the terms of Eq. (\ref{eq:exactMSD}) go towards a constant except the one for $n=0$. Here we have $\tau_0\to \infty$ as $k\to 0$. Expanding this term to first order in $t/\tau_0$ we find the asymptotic behavior at long times to be
\begin{equation}
\langle \delta x_\mT^2(t)\rangle \sim \frac{2 D t}{N}\; ,\quad t\gg N^2\xi/\kappa
\end{equation}
The diffusion is reduced by a factor $1/N$ because we are observing the collective motion of all $N$ particles in this limit. 
To see what happens in the intermediate regime $\xi/\kappa\ll t\ll N^2\xi/\kappa$ we can take $N\to \infty$ before taking $t\to\infty$. In this limit we can do sums of the form
\begin{equation}
g(t)=\frac{1}{N}\sum_n\tau_n\left(1-e^{-t/\tau_n}\right)\label{eq:gdef}
\end{equation}
by differentiating
\begin{equation}
g'(t)=\frac{1}{N}\sum_n e^{-t/\tau_n}
\end{equation}
and then replace $u_n\to u$ and $N^{-1}\sum_n\to (2\pi)^{-1}\int du$. Also, since $t\gg \xi/\kappa$ we only get contributions from terms with $n\ll N$. Thus we can approximate $\tau_n^{-1}\sim \kappa u^2/\xi$ and take the limits of the $u$ integration to infinity. We then get a Gaussian integral which can be evaluated to
\begin{equation}
g'(t)\sim \frac{1}{2\pi}\int_{-\infty}^\infty du\, e^{-\kappa t u^2/\xi}=\sqrt{\frac{\xi}{4\pi\kappa t}}
\end{equation}
Integrating using $g(0)=0$ we then find
\begin{equation}
g(t)\sim \sqrt{\frac{\xi t}{\pi\kappa}}\label{eq:gres}
\end{equation}
Applying this on Eq. (\ref{eq:exactMSD}) we get for $\xi/\kappa\ll t\ll N^2\xi/\kappa$
\begin{equation}
\langle \delta x_\mT^2(t)\rangle \sim \frac{2 k_B T}{\xi}\left[1+\left(1-\frac{1}{\sqrt{2}}\right)\frac{T_{\rm chain}-T}{T}\right]\sqrt{\frac{ \xi t}{\pi \kappa}}
\end{equation}
which is the same as Eq. (\ref{eq:effMSD}). An intermediate step in this calculation was to write
\begin{equation}
(1-e^{-t/\tau})^2=2(1- e^{-t/\tau_n})-(1-e^{-2 t/\tau_n})
\end{equation}
and then also use Eqs. (\ref{eq:gdef}) and (\ref{eq:gres}) with $2 t$ instead of $t$.

Regarding the thermodynamic quantities, then in the absence of external force we have an average 
internal energy which is
\begin{equation}
\langle U_{f=0} \rangle=\frac{1}{2}k_B T \sum_n \left[1-\left(1-
\frac{T_{\rm chain}}{T}\right) e^{-2 t/\tau_n}\right].
\end{equation}
It is interesting to note that in the limit $t \rightarrow 0$,  this equation
gives an internal energy per particle $k_B T_{\rm chain}/2$, which confirms the
interpretation of $T_{\rm chain}$ as a measure of the elastic energy stored in the 
chain at $t=0$. The same property of $T_{\rm chain}$ also holds in the continuous limit, and
is most easily seen in Eq.~\ref{equipart}, which can be understood as
a statement of the equipartition of the elastic energy at $t=0$ for a mode of 
wavevector $q$.  As expected in the limit $t \rightarrow \infty$, the internal
energy per particle is instead $k_B T/2$.

Let us now turn to average stochastic entropy of the system that can be found from the corresponding Gaussian 
probability distribution $P$ as detailed in Appendix \ref{Appendix_A} \cite{Seifert2012} 
\begin{align}
\langle S\rangle = -k_B\langle \ln P\rangle = \frac{N k_B}{2}+\frac{k_B}{2}\sum_n \ln\left(2\pi {\tilde V}_{nn}(t)\right)
\end{align}
Note that it does not affect this result whether we take the probability distribution of the $y_n$ or the ${\tilde y}_n$, since the transformation between them is unitary.

We can compare the above thermodynamic quantities with the results in the continuum limit by taking $N\to \infty$ and assume $t\gg \xi/\kappa$ using the same techniques as for the MSD calculation. Doing this we find
\begin{equation}
\frac{\langle U_{f=0} \rangle}{N}\sim \frac{1}{2}k_B T \left[1-\left(1-\frac{T_{\rm chain}}{T}\right) \sqrt{\frac{\xi}{8\pi\kappa t}}\right],
\end{equation}
which agrees with the result of the continuum calculation in Eq. (\ref{eq:uaverage}), and
\begin{align}
\frac{\langle S\rangle}{N} \sim \frac{k_B}{2}\left[1+\ln\frac{2\pi k_B T}{\kappa} 
+\sqrt{\frac{\xi}{8\pi\kappa t}}\left(\frac{T_{\rm chain}}{T}-1\right)\right].
\end{align}

\section{The effective non-Markovian description of one tagged particle}\label{sec:IV}
We will now focus on formulating this problem from the point of view of the tagged particle. 
To obtain the effective equation for the tagged particle, one should 
integrate out the motion of the other particles as shown in Ref. \cite{lizana10}.
First we Fourier invert Eq. (\ref{eqofmotion}), and then specify $n=0$ in the result. This gives
the following equation for $x_\mT(s)=x(n=0,s)$:
\begin{equation}
\gamma(s) [ s x_\mT(s) - x_\mT(0) ] = {\bar \eta}^{\rm eff}_\mT(s) + f_\mT(s),
\label{gen Langevin Laplace}
\end{equation}
where $\gamma(s)=\sqrt{4\kappa \xi/s}$ and the effective noise ${\bar \eta}^{\rm eff}_\mT$ entering the equation 
now depends on the original noise $\eta(n,t)$ and on the initial conditions:
\begin{equation}
{\bar \eta}^{\rm eff}_\mT(s)={\bar \eta}(s)+{\bar \eta}^{\rm init}(s)
\end{equation}
with
\begin{align}
{\bar \eta}(s)=&\int dn\, \exp(-\sqrt{\xi s/\kappa}|n|)\eta(n,s)\\
{\bar \eta}^{\rm init}(s)=&\xi\int dn\, \exp(-\sqrt{\xi s/\kappa}|n|)[ {x}(n,t=0)-x_\mT(0)]
\end{align}
After Laplace inversion, the effective equation of motion for the tagged particle
 becomes a generalized Langevin equation
\begin{equation}
\int_0^t dt' \gamma(t-t')\frac{d x_\mT(t')}{d t'}={\bar \eta}^{\rm eff}_\mT(t)+f_\mT(t),
\label{gen Langevin}
\end{equation}
with kernel $\gamma(t)=\sqrt{4\xi \kappa/(\pi t)}$.
A calculation of the noise correlations gives \cite{lizana10}
\begin{align}
&\left<{\bar \eta}^{\rm init}(t){\bar \eta}^{\rm init}(t')\right>=k_B T_{\rm chain}\gamma(t+t'),\\
&\left<{\bar \eta}(t){\bar \eta}(t')\right>=k_B T[\gamma(|t-t'|)-\gamma(t+t')].
\end{align}
with no correlation between the two: $\left<{\bar \eta}^{\rm init}(t){\bar \eta}(t')\right>=0$.
Thus we have for the effective noise that
\begin{equation}
\left<{\bar \eta}^{\rm eff}_\mT(t){\bar \eta}^{\rm eff}_\mT(t')\right>=k_BT\gamma(|t-t'|)+k_B(T_{\rm chain}-T)\gamma(t+t').
\label{var_eta_eff}
\end{equation}
Note the loss of translational time invariance in the noise variance when $T \ne T_{\rm chain}$.

\subsection{First law of thermodynamics}
In the effective description in which only the coordinate of the 
tagged particle are kept, the system of interest is
the tagged particle and everything else is part of the
environment. Thus, this system has no internal energy. 
This can be seen by realizing that the stochastic work is still given by 
by the same expression as in Eq.~(\ref{work}), namely:
\begin{equation}
W =\int_0^t d t'\, f_\mT(t') \dot{x}_\mT(t'), 
\end{equation}
while the heat is
\begin{align}
Q_\mT  &= \int_0^t d t'\, [  -{\bar \eta}^{\rm eff}_\mT(t') + \int_0^{t'} dt'' 
\gamma(t'-t'') \dot{x}_\mT(t'') ] \dot{x}_\mT(t'),
\end{align}
so that the first law at the trajectory level indeed reads $W-Q_\mT=0$.
In view of Eqs.~(\ref{internal energy})-(\ref{heat}), we obviously have that 
$Q_\mT=Q + \Delta U$, in other words, what is called heat at the tagged 
particle level can in fact be split into heat exchanged with the heat bath surrounding the chain
and elastic energy stored into the chain.
However, such a decomposition is not accessible in the effective description 
from the first law and could only show up in the second law.
As expected if energy gets stored in the chain ($\Delta U>0$), this appears
for the tagged particle as an increased dissipation (a positive contribution 
to $Q_\mT$).

\subsection{Second law of thermodynamics}\label{subsec:sec_law}
In this section, we follow a path integral formulation introduced in Ref. \cite{Ohkuma2007_vol2007} 
and also used in Ref. \cite{Aron2010} 
to evaluate the entropy production for our generalized Langevin, namely Eq. (\ref{gen Langevin}).
The path probability of a trajectory $\{ x_\mT \}_0^T$ which starts at $x_\mT(0)$ at time 0 and ends at $x_\mT(t_f)$ 
at time $t_f$, $P[x_\mT(t_f) | x_\mT(0)]$, is related to the probability of the noise history $P[{\bar \eta}^{\rm eff}_\mT(..)]$ by
\be
P[x_\mT(t_f) | x_\mT(0)] \mathcal{D}x_\mT = P[{\bar \eta}^{\rm eff}_\mT(..)] \mathcal{D} {\bar \eta}^{\rm eff}_\mT.
\ee 
Since the noise ${\bar \eta}^{\rm eff}_\mT$ is Gaussian and colored, it is such that for any function $z(t)$,
\bea
& & \langle \exp \left( i \int dt\, z(t) {\bar \eta}^{\rm eff}_\mT(t) \right) \rangle = \nonumber \\
& & \exp \left( - \frac{1}{2} \int dt dt' z(t) z(t') v(t,t') \right),
\eea
where $v(t,t')$ is the covariance of the noise ${\bar \eta}^{\rm eff}_\mT$ given 
in general by Eq. (\ref{var_eta_eff}). 
Therefore, the probability of the noise history can be written as
\be
P[{\bar \eta}^{\rm eff}_\mT(..)] \simeq \exp \left( -\frac{1}{2} \int_0^{t_f} dt \int_0^{t_f} dt' 
{\bar \eta}^{\rm eff}_\mT(t) G(t,t') {\bar \eta}^{\rm eff}_\mT(t') \right),
\label{noise-history}
\ee
where $G$ is the inverse of the function $v$ in the interval $[0,t_f]$ in the following sense
\be
\int_0^{t_f} dt' G(t,t') v(t',t'') = \delta(t-t''),
\label{inverseG}
\ee
for $0 \le t,t'' \le t_f$.
Note that, the analytical calculation of $G$ is a difficult task since $v(t',t'')$  
is in general not a function of $t'-t''$, and Eq.~(\ref{inverseG}) does not have the form of a convolution product.

From the generalized Langevin equation Eq.~(\ref{gen Langevin}) and Eq.~(\ref{noise-history}), one obtains
\bea
\label{path integral}
P [x_\mT(t_f) | x_\mT(0)]  &= & J \exp \left( -\frac{1}{2} \int_0^{t_f} dt \int_0^{t_f} dt' G(t,t') \right. \nonumber \\
& & \left[ \int_0^t ds\, \gamma(t-s) \dot{x}_\mT(s) - f_\mT(t) \right]   \\
& & \left. \left[ \int_0^{t'} ds' \gamma(t'-s') \dot{x}_\mT(s') - f_\mT(t') \right]  \right), \nonumber
\eea 
where $J$ is the Jacobian $J=\det \delta {\bar \eta}^{\rm eff}_\mT(t)/ \delta x_\mT(t')$.
The explicit evaluation of this Jacobian will not be needed for the calculation we are interested 
in because it is independent of $x_\mT$.

Let us first discuss the case where $T=T_{\rm chain}$, where there is a proper heat bath, 
which means that its internal degrees of freedom are equilibrated initially and the 
standard framework of stochastic thermodynamics applies. 
In this case, the medium entropy $S_m$ will be related to 
the heat $Q_\mT$ introduced above as shown below. To see this, let us define $\Delta S_m$ by
\be
\label{def-Sm}
\Delta S_m =  k_B \ln \frac{P [x_\mT(t_f) | x_\mT(0)]}{P_B [\hat{x}_\mT(t_f) | \hat{x}_\mT(0)]},
\ee 
where the index $B$ refers to a backward process, which is constructed from the forward
process described by Eq.~(\ref{path integral}) and involves a time-reversal
of the trajectories and protocol denoted by 
$\hat{x}_\mT(t)=x_\mT(t_f -t)$ and $\hat{f}_\mT(t)=f_\mT(t_f-t)$.
In the evaluation of $S_m$, only terms which are antisymmetric under time reversal
remain. In the present case such terms are linear in $\dot{x}_\mT$, since
that quantity changes sign under time reversal symmetry. 
Using the property that $G(t,t')=G(t_f-t,t_f-t')$ one
finds using Eqs. (\ref{path integral})-(\ref{def-Sm}), 
\be
\Delta S_m =  k_B \int_0^{t_f} dt \int_0^{t_f} dt'  \int_0^{t_f} ds \dot{x}_\mT(s) \gamma(|t-s|) G(t,t') f_\mT(t').
\ee
Note the appearance of the absolute value in the $\gamma(|t-s|)$ imposing the positivity of the argument.
Using the definition of $G(t,t')$, namely Eq.~(\ref{inverseG}), \chg{and the expression for the covariance $v(t,t')$ of the effective noise, Eq. (\ref{var_eta_eff}), in equilibrium (that is $T_{\rm chain}=T$)} one can perform the integral over $t$, which gives a delta function. Therefore, one obtains
\be
\Delta S_m =  \frac{1}{T} \int_0^{t_f} dt' \dot{x}_\mT(t') f_\mT(t')= \frac{W}{T}=\frac{Q}{T},
\label{Q-Sm}
\ee
which has the same form as that derived by Crooks for Markovian dynamics.

In addition, by defining the system entropy as the difference
in Shannon entropy constructed from the initial and final probability distributions \cite{Ohkuma2007_vol2007}:
\be
\Delta S=-k_B \ln \frac{p(\hat{x}_\mT(0))}{p(x_\mT(0))},
\label{S_sys}
\ee
one obtains a proper form for the total entropy production: $\Delta S_{\rm tot}=\Delta S 
+ \Delta S_m=k_B \ln P[x_\mT(t_f)]/P_B[\hat{x}_\mT(t_f)]$, which
satisfies by construction the integral fluctuation theorem
$\langle e^{-\Delta S_{tot}/k_B} \rangle =1$. The second law of thermodynamics namely $\langle \Delta S_{\rm tot} \rangle \ge 0$
follows from this theorem. 

One may wonder whether in addition to this integral fluctuation theorem, a detailed integral theorem for the entropy production exists. 
Specific conditions are in fact needed to guarantee such a result as explained in Ref.~\cite{Verley2012a}. In the present problem,
 these conditions on the initial and final probability distributions are not met, since the dynamics 
does not reach a steady state and the driving may not be periodic. 
 
When $T \neq T_{\rm chain}$, we loose the notion of heat bath in the sense that this heat bath now contains
degrees of freedom which are not at equilibrium initially. 
\chg{Another way this could happen 
is if the chain is driven out of equilibrium by the application of  
an external force $f_\mT(t)$ that does not vanish for $t<0$.} 
As a result, if one still defines $S_m$ from Eq.~(\ref{def-Sm}),
then one will find that $\Delta S_m$ is not directly equal to $Q/T$, there is an extra contribution proportional to $T-T_{\rm chain}$,
which can be written formally as an integral involving $G(t,t'), \dot{x}_\mT$ and $\gamma(t)$. 
The explicit expression is involved and is not reported here since it is difficult to evaluate it given the lack of an  
explicit form of the inverse of the variance of the noise, $G(t,t')$ defined in Eq.~(\ref{inverseG}).
Another possible approach which avoids altogether such a correction consists in postulating that $\Delta S_m=Q/T$ 
even when $T \neq T_{\rm chain}$,
 and introducing an ad-hoc modification of the dynamics 
for the backward process in Eq.~(\ref{def-Sm}) to satisfy this. 
Note that this amounts to redefining entropy production as 
the breaking of a more general symmetry different from time-reversal.

\chg{When the system is in equilibrium initially, $T = T_{\rm chain}$, none of these
issues arise since the covariance of the effective noise satisfies Eq. (\ref{var_eta_eff}) and therefore Eq.~(\ref{Q-Sm}) holds. 
Note however, even in this case, the special role played by the origin at time $t=0$, 
due to the non-Markovian nature of the problem.
As a result, if an external force has driven the system out of equilibrium between time $t=0$ and a later time $t=t_0$ then it could be that $\langle S(t_0+\Delta t)\rangle-\langle S(t_0)\rangle$ can become negative.
Therefore, in the limit $\Delta t\to 0$, it is possible that the entropy production rate
could become negative transiently despite the fact that the average entropy
production between time $0$ and time $t$ is always positive. 
In fact, we will see in the next
subsection and in Appendix \ref{Appendix_B} examples where the entropy production rate is temporarily negative.}


\subsection{A negative dissipation rate for the coarse-grained dynamics of the tagged particle}
As we have seen previously the work done by the external force 
on the tagged particle equals the  
heat which is transferred into the surrounding environment, {i.e.}, either 
the chain or the fluid. This surrounding environment  
acts as a visco-elastic medium for the tagged particle, because part of the energy 
which is apparently `dissipated' can be returned to the particle. To illustrate this point, we apply an 
oscillating external force on the tagged particle starting from time $t=0$
\begin{equation}
f_\mT(t) = F_0 \cos(\omega_0 t), 
\label{oscillating force}
\end{equation}
and we introduce the frequency dependent mobility $\mu(\omega)$ from
the response of the velocity of the tagged particle $v_\mT$ to the force $f_\mT$:
$\langle v_\mT ( \omega) \rangle= \mu(\omega)  f_\mT(\omega)$,
where the Fourier transform in time of a function $f(t)$ is simply denoted $f(\omega)$.
By Fourier transforming Eq. (\ref{Langevin}) in space and time, isolating position and inverting the Fourier transform at $n=0$, one obtains
\be
\mu ( \omega )=\int \frac{dq}{2 \pi} \frac{-i \omega}{\kappa q^2 - i \omega \xi} = \frac{1}{4} \sqrt{ \frac{2 \omega}{\kappa \xi}} (1-i).
\ee
This then yields
\begin{equation}
\langle x_\mT(t) \rangle
\sim\frac{F_0} {2\sqrt{\omega_0\kappa \xi}} \cos\left( \omega_0 t-\pi/4\right).
\end{equation}
The $\sim$ here indicates that the expression is valid asymptotically for large $t$, since the Fourier transform in time implies that the effect of initial conditions has been pushed back to $t=-\infty$. A more precise calculation
including the motion of the tagged particle at short time leads to:
\begin{align}
\langle x_\mT(t) \rangle 
&=\frac{F_0}{2\sqrt{\pi\kappa\xi}}\int_0^t d t'\; \cos(\omega_0(t-t'))\frac{1}{\sqrt{t'}}
\nonumber \\
&=\Re \left[ F_0 \frac{e^{i\omega_0 t} \, {\rm erf}(\sqrt{i\omega_0t})}{2\sqrt{i\omega_0\xi\kappa}} \right].
\end{align}

\begin{figure}
\includegraphics[width=8cm]{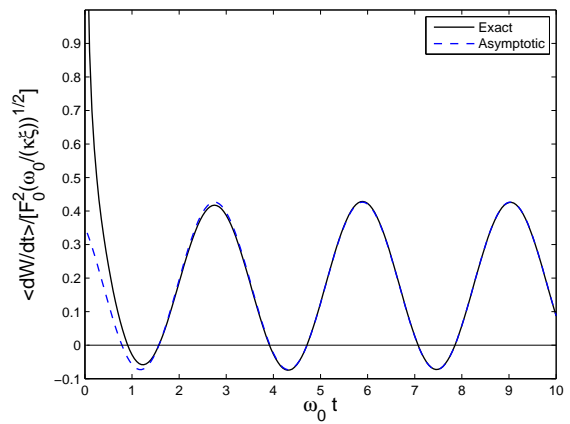}
\caption{Plot of the exact (solid line) and asymptotic (Eq. (\ref{eq:Wdot}), dashed line) 
$\langle{\dot W}\rangle/(F_0^2\sqrt{\omega_0}/(\sqrt{\kappa\xi}))$ as a function of $\omega_0 t$ for a harmonically oscillating force.} 
\label{fig:Wdotplot}
\end{figure}

In any case, the rate of the average work done on the particle behaves at large times as
\begin{equation}
\langle {\dot W}\rangle \sim\frac{F_0^2}{8\sqrt{\xi\kappa\omega_0}}\left[\sqrt{2}\omega_0-2\omega_0\sin(2\omega_0 t-\pi/4)\right]. \label{eq:Wdot}
\end{equation}
It is interesting to note that this quantity can be temporarily negative as shown in Fig. (\ref{fig:Wdotplot}). 

In order to relate this to the dissipation, we go back to the definition of the average entropy production
 rate, which quantifies the dissipation, namely $\langle \Delta S_{\rm tot} \rangle =\langle \Delta S \rangle + 
\langle \Delta S_m \rangle = 
 \langle \Delta S \rangle + \langle W \rangle /T$. We recall that the system entropy is 
a state function, therefore $\Delta S(t)= S(t) - S(0)$,
and $S$ itself should be evaluated from Eq.~(\ref{S_sys}) as explained in Appendix \ref{Appendix_A}. 
Thus, one finds that the variation of the average system 
entropy between the initial time and time $t$ is 
\be
\langle \Delta S (t) \rangle =\frac{k_B}{2}  \ln \frac{\sigma_t^2}{\sigma_0^2}.
\ee
in terms of the variance of the position of the tagged particle, $\sigma_t^2$.
Note the presence of the variance of the initial condition, which is zero for a deterministic initial condition,
implying a diverging contribution. 
The rate of variation of the system entropy, however, is in this case well defined, since this 
 diverging factor disappears in the time derivative. Thus, the average rate of change of the system entropy is in that case 
\be
\langle \dot{S}_t \rangle = \langle \Delta \dot{S} (t) \rangle = k_B \frac{ \dot{\sigma}_t }{\sigma_t} = \frac{k_B}{4 t},
\label{S_dot}
\ee
where the last expression is obtained from Eq.~(\ref{eq:effMSD}) using that for a deterministic initial condition we have $\sigma_t^2=\langle \delta x_\mT^2(t)\rangle$.

\begin{figure}
\includegraphics[width=8cm]{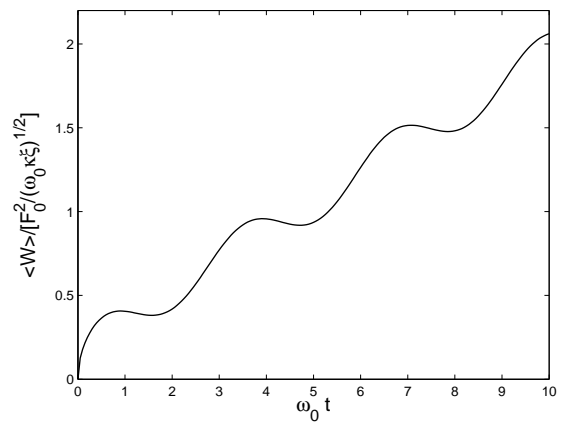}
\caption{Plot of the exact $\langle W\rangle/(F_0^2/(\sqrt{\omega_0\kappa\xi}))$ as a function of $\omega_0 t$ 
(the integral of the solid line in Fig. \ref{fig:Wdotplot}).} \label{fig:Wplot}
\end{figure}

The total amount of dissipation is the sum of a part due to the work and another due to the system entropy: 
$\langle \Delta S_{\rm tot}(t) \rangle = \langle W(t) \rangle /T + \langle \Delta S (t)  \rangle$.
We note that $\langle W(t) \rangle$ is always positive, at all times, as shown on Fig. \ref{fig:Wplot}.

Let us now discuss the average dissipation rate, namely $\langle \Delta \dot{S}_{\rm tot}(t) \rangle$, which contains 
two parts $\langle \dot{W}(t) \rangle /T$ and $\langle \Delta \dot{S} (t)  \rangle$.
From Eqs.~(\ref{eq:Wdot})-(\ref{S_dot}), it is apparent that the contribution due to the work can always be made to dominate 
over the contribution of the system entropy at sufficiently long times or for sufficiently large external force. 
As a result, we have temporarily a negative total dissipation rate $\langle \Delta \dot{S}_{\rm tot}(t) \rangle$. This is not in contradiction
with the second law which only imposes that  $\langle \Delta S_{\rm tot}(t) \rangle \ge 0$, because $\Delta S_{\rm tot}(t)$ is not a  
state variable but a trajectory dependent quantity.
 
At the level of the full chain however, we can calculate the internal energy of the system 
\begin{equation}
\langle \delta U_f\rangle \sim \frac{F_0^2}{8\sqrt{\xi\kappa\omega_0}}\left[\frac{1}{\sqrt{2}} +\frac{1}{2}\cos(2\omega_0 t-\pi/4)\right] 
\end{equation}
Differentiating this and subtracting from the work, we find at long times
\begin{equation}
\langle {\dot W}-\delta{\dot U}_f\rangle \sim\frac{F_0^2}{8\sqrt{\xi\kappa\omega_0}}\left[\sqrt{2}\omega_0-\omega_0\sin(2\omega_0 t-\pi/4)\right].
\end{equation}
Since the system entropy is negligible at long times, this represents the
dissipation rate in the full Markovian description, and it is never negative.

It is interesting to discuss our result in light of recent formulations of entropy production 
for a coarse-grained dynamics such as \cite{Esposito2012_vol85}
and \cite{Bo2014}. In the former reference, entropy production is coarse-grained within 
a master equation formulation starting from an initial Markovian description. 
The obtained coarse-grained entropy production is decomposed into three contributions, where the first contribution
denoted $\langle \Delta \dot{S}^{(1)}_i \rangle$, 
represents the entropy production evaluated using coarse-grained variables. 
It is shown there that this term is always positive in apparent contradiction with 
our result that $\langle \Delta \dot{S}_{\rm tot}(t) \rangle$ may be negative.
A similar result is derived in a more general setting 
including diffusive processes and time dependent driving in Ref.~\cite{Bo2014}.
On closer inspection however, it appears that the results of these references may not contradict 
ours. For instance, the expression of $\langle \Delta \dot{S}^{(1)}_i \rangle$ in \cite{Esposito2012_vol85} 
is defined in terms of transition rates which can only be obtained by first coarse-graining the Markovian description.
This is not the procedure which we have followed here to define $\langle \Delta \dot{S}_{\rm tot}(t) \rangle$.
We indeed defined this quantity directly from the dynamics generated by the generalized Langevin equation 
with no reference to an underlying Markovian level. Given such a difference in both formulations, 
we believe that our finding of a negative dissipation rate at a coarse-grained level 
is not in contradiction with the studies mentioned above.  

To summarize results of this section, whether at a Markovian or non-Markovian level, 
the average entropy production (or dissipation) 
is always positive as imposed by the second law. 
However, the average dissipation rate in the non-Markovian level can be negative, 
while in the corresponding Markovian level, it will remain positive. 
We attribute this difference to the memory present at the non-Markovian level. 
To confirm that this effect is present independently of many details of the model, in particular the system size, we
analyze in Appendix \ref{Appendix_B} a similar problem for just two particles. These two particles are coupled harmonically and 
we focus on the reduced non-Markovian description obtained by integrating out the position of one particle,
with a time dependent force being applied to the remaining particle.
This setting has similarities with the one of Ref.~\cite{Crisanti2012}, where however no
time dependent external force was considered.
In this simpler two particle model, we confirm that there as well it is possible 
to observe a negative dissipation rate when evaluating the entropy production 
for the coarse-grained variable.  

\subsection{Negative dissipation rate for the tagged particle 
within the stochastic entropy formulation}\label{subsec:stochent}

In this section, we explain how a negative dissipation rate can occur within 
the classic formulation of the second law at the trajectory level \cite{Seifert2005a}
in the specific case of non-Markovian dynamics.
 
As done in the previous section, we define the stochastic entropy of the tagged particle as 
\be
S_t=-k_B \ln p_t(x_\mT(t)),
\label{Stoch_entropy}
\ee
so that the system entropy introduced in Eq.~(\ref{S_sys}) is the difference of stochastic entropy between 
the time $t$ and time 0.
From this definition, taking a time derivative, one obtains
\be 
{\dot S}_t=-k_B \frac{{\dot p}_t(x_\mT(t))}{p_t(x_\mT(t))}  - k_B {\dot x}_\mT(t) \partial_{x_\mT} \ln p_t(x_\mT(t)),
\label{s_dot}
\ee
As mentioned earlier in the context of the work distribution, the probability density of $x_\mT$ is Gaussian: 
\be
p_t(x_\mT)=\frac{1}{\sqrt{2 \pi \sigma_t^2}} e^{-\left( x_\mT - \langle x_\mT \rangle \right)^2/(2 \sigma_t^2)},
\label{gaussian_pdf}
\ee
where $\sigma_t^2=\langle \delta x_\mT^2 \rangle$ is the variance of $x_\mT$ introduced earlier. It is a simple matter to check that this Gaussian density satisfies the conservation law
\be
\partial_t p_t(x)+\partial_x[p_t(x)\nu_t(x)]=0,
\ee
where we have introduced the local current or drift velocity
\chg{
\be
\nu_t(x)=\langle {\dot x}_\mT(t) \rangle - \Dt  \partial_{x} \ln p_t(x),
\ee
The quantity $\Dt =\sigma_t  {\dot \sigma}_t$ is a positive 
time dependent diffusion coefficient.}
With these notations we can rewrite Eq.~(\ref{s_dot}) as
\be
{\dot S}_t= - k_B (\partial_t \ln p_t)(x_\mT(t)) - k_B{\dot x}_\mT(t) \frac{  \langle {\dot x}_\mT \rangle - \nu_t (x_\mT(t))}{\Dt} .
\label{s_dot2}
\ee
We note that the local current $\nu_t$ is well defined here due to the Gaussian property of the probability 
distribution $p_t(x_\mT)$, despite the fact that this probability does not satisfy a standard Fokker-Planck equation
as in the Markovian case. We now take an average with respect to the trajectories. 
Due to the probability conservation, the first term in Eq.~(\ref{s_dot2}) becomes zero upon averaging. We are thus left with
 \be
\langle {\dot S}_t \rangle = -\frac{k_B \langle  {\dot x}_\mT(t) \rangle^2}{\Dt} 
+ \frac{k_B \langle {\dot x}_\mT(t) \cdot \nu_t(x_\mT(t)) \rangle}{\Dt}, 
\label{s_dot_av}
\ee
where $\cdot$ denotes a Stratonovich product.
The second term in the above equation can be written 
\be
\langle {\dot S}_i \rangle = \frac{k_B}{\Dt} \langle {\dot x}_\mT(t) \cdot \nu_t(x_\mT(t)) \rangle= 
\frac{k_B}{\Dt} \int dx\, \nu_t(x)^2 p_t(x),
\label{S_i}
\ee
which is a positive quantity and is the equivalent of the entropy production for the Markovian case \cite{Seifert2005a}.

In contrast to the Markovian case however, the first term in Eq.~(\ref{s_dot_av}) is distinct from the quantity defined earlier as 
$\langle {\dot S}_m \rangle$ in Eq.~(\ref{Q-Sm}), which is related to the rates of work and heat when $T=T_{chain}$ by
$\langle {\dot S}_m \rangle=\langle {\dot Q} \rangle/T=\langle {\dot W} \rangle/T={f_\mT(t)} \langle  {\dot x}_\mT(t) \rangle/T$. 
To pinpoint this difference, we
add and subtract this quantity to the right hand side of Eq.~(\ref{s_dot_av}), which now reads
\be
\langle {\dot S}_t \rangle = \left( -\frac{k_B \langle  {\dot x}_\mT(t) \rangle}{\Dt} +\frac{ f_\mT(t) }{T} 
\right) \langle  {\dot x}_\mT(t) \rangle
- \langle {\dot S}_m \rangle + \langle {\dot S}_i \rangle.
\ee
This equation may be rewritten equivalently in terms of the entropy production $S_{tot}$, which we defined earlier as the sum of $S_t+S_m$, as
\be
\langle {\dot S}_{tot} \rangle = \langle {\dot S}_{nM} \rangle + \langle {\dot S}_i \rangle,
\label{2nd law NM}
\ee
where we have introduced $\langle {\dot S}_{nM} \rangle$ defined as
\be
\langle {\dot S}_{nM} \rangle=\left( -\frac{k_B \langle  {\dot x}_\mT(t) \rangle}{\Dt} 
+ \frac{ f_\mT(t) }{T} \right) \langle  {\dot x}_\mT(t) \rangle.
\label{S_nm}
\ee
We call this new quantity $\langle {\dot S}_{nM} \rangle$ a non-Markovian contribution to the entropy production, since it is 
absent in the Markovian case. It is important to note that this term can be positive or negative. 
When it is sufficiently negative, it can overcome the positive term $\langle {\dot S}_i \rangle$
with the result that $\langle {\dot S}_{tot} \rangle$ can be negative. We recall that $\langle {\dot S}_{tot} \rangle$ represents 
the coarse-grained entropy production of the tagged particle.

To investigate this non-Markovian contribution, we go back to the generalized Langevin equation for the tagged particle motion. 
From the equation of motion in Laplace space namely Eq.~(\ref{gen Langevin Laplace}), we divide by $s$ and  
we use the convolution theorem to obtain
\be
x_\mT(t)=x_\mT(0) + \frac{1}{4 \xi \kappa} [ \int_0^t dt' \left( {\bar \eta}^{\rm eff}_\mT(t')+f_\mT(t') \right) \gamma(t-t') ].
\ee
From this, after taking a time derivative and an average, and then integrating by parts, one obtains
\be
\langle {\dot x}_\mT(t) \rangle = f_\mT(t=0^+) \gamma(t) + \int_{0^+}^t dt' {\dot f}_\mT(t') \gamma(t-t').
\ee

For the particular case of a constant force turned on at $t=0$, i.e., $f_\mT(t)=F \Theta(t)$, 
one indeed obtains from this $\langle {\dot x}_\mT(t) \rangle=F/2 \sqrt{\kappa \xi \pi t}$.
When reporting this into Eq.~(\ref{S_nm}), one finds $\langle {\dot S}_{nM} \rangle=0$. 
This means that $\langle {\dot S}_{tot} \rangle \ge 0$ which is expected since in this case
$\langle {\dot W} \rangle \ge 0$.

\begin{figure}
\includegraphics[width=8cm]{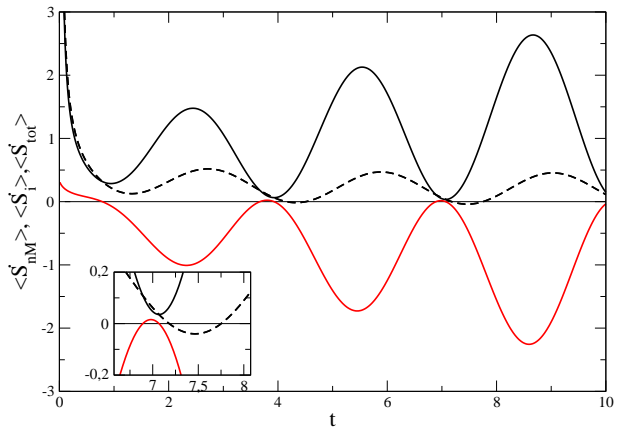}
\caption{Plot of $\langle {\dot S}_{nM} \rangle$ (red solid curve mainly negative), $\langle {\dot S}_i \rangle$ (black solid 
curve only positive) and $\langle {\dot S}_{tot} \rangle$ (dashed line) versus time $t$,
for a harmonically oscillating force. For simplicity we have assumed $k_B=\kappa=\xi=F_0=\omega_0=1$. Inset: zoom of 
the main figure for $t=6.5$ and $t=8$, which makes clear that $\langle {\dot S}_{tot} \rangle$ can become negative.} \label{fig:Snm}
\end{figure}

In contrast to this, in the case of the oscillating force considered in Eq.~(\ref{oscillating force}), 
one finds a non-zero mainly negative $\langle {\dot S}_{nM} \rangle$ which can
indeed transiently overcome the positive term $\langle {\dot S}_i \rangle$.
An explicit evaluation of $\langle {\dot S}_i \rangle$ using Eq.~(\ref{S_i}) confirms that 
\be
\langle {\dot S}_i \rangle = \frac{k_B}{4t} + \frac{k_B \langle {\dot x}_\mT(t) \rangle^2}{\Dt}.
\ee
In this expression, one recognizes in the first term $k_B/(4t)$ the average rate of system entropy $\langle {\dot S}_t \rangle$
already obtained in Eq.~(\ref{S_dot}), while the second term corresponds to the form which $\langle {\dot S}_m \rangle$ would take 
in the Markovian case.
This equation also implies that the non-Markovian contribution is indeed given by Eq.~(\ref{S_nm}).
In order to illustrate the relative contribution of the various terms in the entropy production
for the case of an oscillating force, we show in Fig. \ref{fig:Snm} a plot of  $\langle {\dot S}_{nM} \rangle$, 
$\langle {\dot S}_i \rangle$ and $\langle {\dot S}_{nM} \rangle$ versus time. On this plot, one can see a large cancellation 
between $\langle {\dot S}_{nM} \rangle$,
which is mainly negative, and $\langle {\dot S}_i \rangle$ which is always positive, so that the result $\langle {\dot S}_{tot} \rangle$
is of smaller amplitude, mainly positive except for small regions, where it can become transiently negative, in 
agreement with what we found in Fig.~\ref{fig:Wdotplot}. 

\subsection{Modified fluctuation-dissipation theorem}
In recent years, many works have been devoted to generalizations of the fluctuation-dissipation theorem for systems 
out of equilibrium \cite{Verley2012}. Let us very briefly illustrate a couple of these approaches for the 
harmonic chain studied using the effective non-Markovian description given by Eq.~(\ref{gen Langevin}). 
One approach is based on an extension of a fluctuation relation called the Hatano-Sasa relation \cite{Hatano2001_vol86} 
 for general out-of-equilibrium systems. This method originally developed for non-equilibrium steady states \cite{Prost2009_vol103}, 
was later extended for arbitrary non-stationary states ruled by Markovian dynamics \cite{Verley2011_vol}.
Under certain stability conditions, the result also holds for non-Markovian systems \cite{Garcia-Garcia2012}. 
The second approach is based on the notion of frenezy, defined as the time-symmetric part of the action for the system of interest
(see \cite{Baiesi2009_vol137} and references therein).

For the first approach, let us introduce the response function of the tagged particle position 
due to the application of a small force $f_\mT$ about a
state with zero force:
\be
R(t,t')= \left. \frac{\delta \langle x_\mT(t) \rangle}{\delta f_\mT(t')} \right|_{f_\mT=0},
\ee
which is non-zero only for $t>t'$. 
In this linear Langevin framework, it is easy to check that this response function should be
independent of the initial condition. Therefore, it should be given by the equilibrium result,
which has time translational invariance. In view of this, we introduce the function  
$\tilde{R}(t-t')=R(t,t')$, and $\tilde{R}(s)$, its Laplace transform, defined as usual by 
\be
\tilde{R}(s)=\int_0^\infty dt\, e^{-st} \tilde{R}(t).
\ee
We also introduce the integrated response function as 
\be
\chi(t,t')=\int_{t'}^t d \tau R(t,\tau).
\ee

From Eq. (\ref{gen Langevin}), the Laplace transform of the response function satisfies 
$\langle x_\mT(s) \rangle=\tilde{R}(s) f(s)$, from 
which we obtain 
$\tilde{R}(s)=1/ (s \gamma(s))=1/\sqrt{4 \kappa \xi s}$. By inverting this Laplace transform and 
using time translational invariance of the 
response function, one obtains:
$R(t,t')=1/\sqrt{4 \kappa \xi \pi (t-t')}$. This corresponds to the integrated response function
\be
\chi(t,0)=\sqrt{ \frac{t}{\kappa \xi \pi} }.
\ee

As shown in Ref. \cite{Verley2011_vol}, a modified fluctuation-dissipation theorem 
valid near any non-equilibrium state (whether it is stationary or not) can be written as
\be
R(t,t')=-\frac{d}{dt'} \langle \partial_{f_\mT} s_{t'}(x_\mT(t'),f_\mT) x_\mT(t) \rangle,
\label{MFDT}
\ee
where $s_t(x_\mT,f_\mT)$ is a trajectory entropy closely related but distinct 
from the stochastic entropy $S_t$ defined in Eq.~(\ref{Stoch_entropy}). 
The quantity $s_t(x_\mT,f_\mT)$ is defined as $s_t(x_\mT,f_\mT)= -\log p_t(x_\mT,f_\mT)$,
in terms of $p_t(x_\mT,f_\mT)$, the probability density of $x_\mT$ for a frozen driving force $f_\mT$.
The derivative inside the correlation function is a normal derivative evaluated at $f_\mT=0$,
while the average $\langle .. \rangle$ is a non-perturbed non-equilibrium average.
 
If we assume that the force $f_\mT(t)$ is turned on at time $t=0$, then we need to focus only on
\be
\chi(t,0)=-\langle \partial_{f_\mT} s_t(x_\mT(t),f_\mT(t)) x_\mT(t) \rangle.
\label{chi}
\ee
We shall use again the property that the probability density of $x_\mT$ is Gaussian, as in 
Eq.~(\ref{gaussian_pdf}) with a variance  $\sigma_t^2=\langle \delta x_\mT^2 \rangle$
which importantly is independent of the applied force.
As a result
\bea
\partial_{f_\mT} s_t(x_\mT(t),f_\mT) &=& \frac{x_\mT(t) - 
\langle x_\mT(t) \rangle}{\sigma_t^2} \partial_{f_\mT} \langle x_\mT(t) \rangle, \nonumber \\
&=& \frac{x_\mT(t) - \langle x_\mT(t) \rangle}{\sigma_t^2} \sqrt{\frac{t}{\pi \kappa \xi}}. 
\eea
Reporting this expression into Eq. (\ref{chi}), one recovers the expected result
\be
\chi(t,0)= \sqrt{\frac{t}{\pi \kappa \xi}},
\ee
independently of the choice of initial conditions since the variance which depends on it cancel in the 
evaluation of $\chi(t)$.

Beyond linear perturbation, the distribution $p_t(x_\mT,f_\mT)$ can be used to construct a work-like
functional $Y=Y(t)$ 
\be
Y=\int_0^t dt' \dot{f}_\mT(t') \partial_{f_\mT} s_{t'}(x_\mT(t'),f_\mT(t')),
\ee
which satisfies a modified Hatano-Sasa relation.
By expanding this fluctuation relation to first order in the perturbation field which is here the force $f_\mT(t)$, 
one obtains the modified fluctuation-dissipation theorem
of Eq. (\ref{MFDT}) \cite{Verley2011_vol}.
In this reference, this theorem was derived under the assumption that the dynamics was Markovian, which is 
not the case here. This supports the view that this formulation 
of the modified fluctuation-dissipation theorem is in fact more general. As explained in Ref.~\cite{Garcia-Garcia2012}, 
such a formulation should hold for arbitrary non-Markovian processes provided they satisfy certain stability conditions.  

Let us now very briefly mention one aspect of the second approach \cite{Baiesi2009_vol137},
which is not based on fluctuation theorems, but on the notion of frenezy defined above 
as the time-symmetric part of the action for the system of interest. 
For a complete presentation of that method and specifically for its application  
to systems with strong memory, similar to the system studied in this paper, we
refer the reader to Ref. \cite{Maes2013}.

In the present situation where the generator of the dynamics is linear, 
one obtains the following expression of the response function 
(still assuming $t>t'$): 
\cite{Baiesi2009_vol137} 
\be
R(t,t')=  \frac{1}{2 k_B T} \left( \frac{d}{dt'} \langle x_\mT(t) x_\mT(t') \rangle - \frac{d}{dt} 
\langle x_\mT(t) x_\mT(t') \rangle \right),
\label{maes form}
\ee 
where the first term on the right hand side corresponds to one half the equilibrium correlation, 
while the second term accounts for the frenetic contribution. This frenetic contribution is equal to 
the first term only at equilibrium, in which case both terms add up to give the standard form
of the fluctuation-dissipation theorem.

Explicit expressions for the unequal time correlation function have been obtained before \cite{leibovich13}.
In the case of an initial condition corresponding to $T_{\rm chain}=0$, in which particles are initially on a 
lattice, the expression is 
\be
\langle x_\mT(t) x_\mT(t') \rangle |_{T_{\rm chain}=0}=\frac{k_B T}{\sqrt{\kappa \xi \pi}} \left( \sqrt{t+t'} - \sqrt{t-t'} \right),
\ee
still assuming $t>t'$.
When using this in Eq.~(\ref{maes form}), one indeed recovers the expected result:
\be
R(t,t')=\frac{1}{\sqrt{4 \kappa \xi \pi (t-t')}}.
\ee

\section{The effective non-Markovian description of two tagged particles} \label{sec:V}

In this section we briefly study the dissipation rate in an effective description 
which includes an internal energy. We could include the possibility of a non-zero internal 
energy by hand by introducing a potential acting on the tagged particle. However, another possibility, 
which is the one we pursue here, is to investigate the dynamics of the relative distance 
between two particles. Thus we focus on the coordinate $x_{\rm dist}(t)=x(n_{\rm a},t)-x(n_{\rm d},t)$ 
between two particles in the chain that could be called for instance an "acceptor'' particle 
($n_{\rm a}$) and a "donor'' particle ($n_{\rm d}$). In \cite{lizana10} it was shown that the equation 
of motion for $x_{\rm dist}$ can be written in the form of a generalized Langevin equation
\begin{equation}
\int_0^t d t'\; {\cal K}(t-t') \frac{d x_{\rm dist}(t')}{d t'}=
\eta_{\rm dist}(t)- {\cal U}'(x_{\rm dist}) + f_{\rm dist}(t),
\label{eq:distGLE}
\end{equation}
where $\eta_{\rm dist}(t)$ is a zero-mean noise, the friction kernel in Laplace space is given by
\begin{equation}
{\cal K} (s)=\frac{\gamma(s)}{2(1-e^{-\sqrt{s\kappa \xi}/ k_{\rm dist}})}
-\frac{k_{\rm dist}}{s},
\end{equation}
and the conservative force is
\begin{equation}
-{\cal U}'(x_{\rm dist})=-k_{\rm dist} \delta x_{\rm dist}(t),
\end{equation}
with $\delta x_{\rm dist}(t)=x_{\rm dist}(t)-(n_{\rm a}-n_{\rm d})/\rho$ and the spring constant
$k_{\rm dist}=\kappa/|n_{\rm a}-n_{\rm d}|$.
The external force $f_{\rm dist}(t)=[f(n_{\rm a},t)-f(n_{\rm d},t)]/2$ is conjugate to $x_{\rm dist}$ in the sense that the total rate of work on the system ${\dot W}_{\rm total}=f(n_{\rm a},t){\dot x}(n_{\rm a},t)+f(n_{\rm b},t){\dot x}(n_{\rm b},t)$ can alternatively be written
\begin{equation}
{\dot W}_{\rm total}=f_{\rm dist}(t){\dot x}_{\rm dist}(t)+f_{\rm cm}(t){\dot x}_{\rm cm}(t)
\end{equation}
where $x_{\rm cm}=[x(n_{\rm a},t)+x(n_{\rm d},t)]/2$ is the center of mass coordinate and $f_{\rm cm}(t)=f(n_{\rm a},t)+f(n_{\rm d},t)$ its conjugate force. The thermodynamics of the two coordinates $x_{\rm dist}$ and $x_{\rm cm}$ decouples and we can study the coordinate $x_{\rm dist}$ independently.

\begin{figure}
\includegraphics[width=8cm]{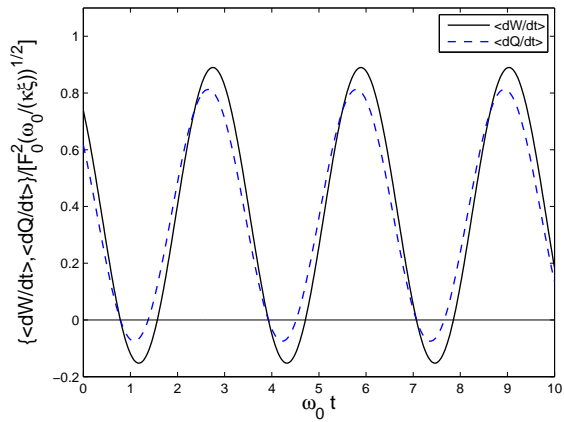}
\caption{Plot of $\langle{\dot W}\rangle/(F_0^2\sqrt{\omega_0}/\sqrt{\kappa\xi})$ (solid line) and $\langle{\dot Q}\rangle/(F_0^2\sqrt{\omega_0}/\sqrt{\kappa\xi})$ (dashed line) as a function of $\omega_0 t$ for a harmonically oscillating force with large frequency: $\omega_0\kappa\xi/k_{\rm dist}^2=20$.} \label{fig:Wdistplot}
\end{figure}

\begin{figure}
\includegraphics[width=8cm]{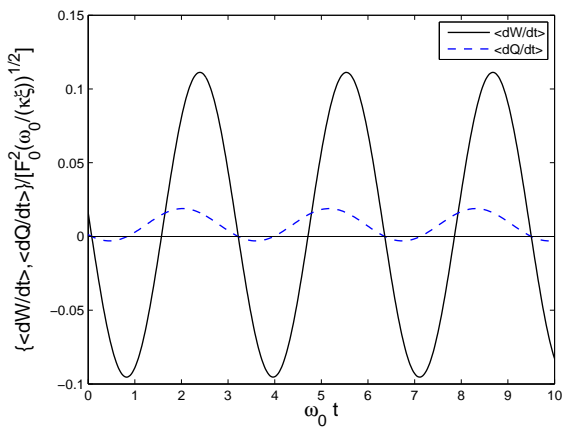}
\caption{Plot of $\langle{\dot W}\rangle/(F_0^2\sqrt{\omega_0}/\sqrt{\kappa\xi})$ (solid line) and $\langle{\dot Q}\rangle/(F_0^2\sqrt{\omega_0}/\sqrt{\kappa\xi})$ (dashed line) as a function of $\omega_0 t$ for a harmonically oscillating force with small frequency: $\omega_0\kappa\xi/k_{\rm dist}^2=1/20$.} \label{fig:Wdistplot2}
\end{figure}

If we look at a harmonically oscillating force $f_{\rm dist}(t)=F_0\cos(\omega_0 t)$ as in Section \ref{sec:IV} we can obtain the behavior of $x_{\rm dist}$ by solving for the mobility in Fourier space $\mu_{\rm dist}(\omega)=f_{\rm dist}(\omega)/\langle{v}_{\rm dist}(\omega)\rangle$ where ${v}_{\rm dist}=d{x}_{\rm dist}/d t$. This can be achieved by shifting the initial time $t=0$ back to $t=-\infty$ to change the lower limit of the integration over the friction kernel to $-\infty$ and then Fourier transforming. The result for the mobility is
\begin{equation}
\mu_{\rm dist}(\omega)=\frac{1-e^{-\sqrt{-i\omega\kappa\xi}/k_{\rm dist}}}{\sqrt{\xi\kappa/(-i\omega)}}
\end{equation}
In this case, two very different limits of the behavior emerges. At large frequencies we have
\begin{equation}
\mu_{\rm dist}\sim \sqrt{\frac{-i\omega}{\kappa\xi}}=\frac{2}{\gamma(s=-i\omega)}\quad [\omega\gg k_{\rm dist}^2/(\kappa\xi)]
\end{equation}
At high frequencies the two particles do not have time to interact, and the mobility becomes twice that of a single tagged particle, with the doubling coming from the factor $1/2$ in the definition of $f_{\rm dist}$. In the opposite limit of small frequencies we get
\begin{equation}
\mu_{\rm dist}\sim \frac{-i\omega}{k_{\rm dist}}\quad [\omega\gg k_{\rm dist}^2/(\kappa\xi)]
\end{equation}
Here we are in the quasi-static limit where the mobility mobility is entirely dominated by the elastic response from the potential between the two particles. If we look at the rate of work and heat exchange we can calculate these as
\begin{align}
\langle {\dot W}_{\rm dist}\rangle & = f_{\rm dist}(t) \langle v_{\rm dist}(t)\rangle\\
\langle {\dot Q}_{\rm dist}\rangle & = \langle {\dot W}_{\rm dist}\rangle - \langle {\dot U}_{\rm dist}\rangle\nonumber\\
& =\langle v_{\rm dist}(t)\rangle \left[f_{\rm dist}(t)-k_{\rm dist}\langle \delta x_{\rm dist}(t)\rangle\right]
\end{align}
where
\begin{align}
\langle v_{\rm dist}(t)\rangle &= {\rm Re}\left[F_0\mu(\omega_0) e^{-i\omega_0 t}\right]\\
\langle \delta x_{\rm dist}(t) \rangle &= {\rm Re}\left[F_0\mu(\omega_0) e^{-i\omega_0 t}/(-i\omega_0)\right]
\end{align}
Here we have assumed that the variance of the position $x_{\rm dist}(t)$ has its stationary value, which exists in this case due to the harmonic
 interaction potential. Thus fluctuations do not contribute to the averages. Using Mathematica we have plotted the rate of work and 
heat exchange for a large frequency in Fig. \ref{fig:Wdistplot} and small frequency in Fig. \ref{fig:Wdistplot2}. Note that at large frequency
 the work is almost entirely dissipated, while in the quasi-static limit the dissipation as heat is becoming vanishingly small. The temporary 
negative rates of heat dissipation persist in both limits, but in the quasi-static limit the overall dissipation becomes vanishingly small 
compared with work and thereby also internal energy.

\section{Conclusion}

We have studied an overdamped harmonic chain from the point of view of stochastic thermodynamics, 
with a particular attention to memory effects which arise when all the degrees of freedom of the chain
have been integrated out except one or two corresponding to one or two tagged particles whose dynamics 
we are interested in tracking. Such a dynamics is non-Markovian, and never reaches a steady state. 
 
We have also introduced the notion of a chain temperature $T_{\rm chain}$, which 
represents the temperature of a heat bath with which the system could equilibrate  
at $t=0$. As we have shown, this temperature measures the elastic energy present in the chain
in the initial state. 
When at time $t>0$, the system is put in contact with a heat bath at temperature $T$, 
the system is driven out of equilibrium and the subsequent application of a force leads to a violation of the Einstein
relation, of the fluctuation-dissipation theorem or of related 
fluctuation theorems like the Crooks relation. The investigation of how this precisely
happens is an active area in the field of stochastic thermodynamics \cite{Maes2013,Maes2013a,Gradenigo2012}.

The case $T_{\rm chain} \ne T$ is also an interesting situation for which the standard formulation
of stochastic thermodynamics fails for the tagged particle, because that particle
is no longer in contact with a proper heat bath, but with a non-equilibrium heat bath,
containing the slow hidden degrees of freedom from the rest of the chain. 
This failure of stochastic thermodynamics manifests itself 
in the fact that the medium entropy denoted $\Delta S_m$ is no
longer simply related to heat, when entropy production is defined as usual from the breaking of 
time reversal symmetry in path probabilities.  
If one instead chooses to postulate that $\Delta S_m$ still represents the heat, then one needs to identify 
a new symmetry breaking of the path probabilities, different from time reversal symmetry, to define 
entropy production.

For the non-Markovian dynamics of the tagged particle, we have also found surprisingly 
that the apparent average dissipation rate can become transiently negative, 
while the same quantity would always remain positive at the Markovian level. 
This negative dissipation rate is allowed by the second law which only enforces the average dissipation to be positive. 
At the trajectory level, we also explained this effect using a decomposition of the stochastic entropy.
We believe the mechanism responsible for a negative apparent dissipation rate to be rather general and therefore, we expect 
that it will occur in many other systems in which memory effects/visco-elasticity are present.
We hope that our study could be useful in triggering more experimental or theoretical studies on these questions.

\section{Acknowledgements}
We would like to thank particularly M. Rosinberg, for a careful reading of this paper, 
and many illuminating discussions and suggestions. 

\appendix
\section{Stochastic system entropy}
\label{Appendix_A}
In this appendix, we calculate the stochastic system entropy 
of some column vector of position(s) $x(t)$ 
for a system of generalized 
linear Langevin equations, and we justify that \chg{the average of} this quantity  
is independent of the external force entering in the Langevin equation:
\begin{equation}
\gamma \ddt x(t) = -A x(t) + f(t) + \eta(t)
\end{equation}
with $\gamma$ being some friction constant, $\ddt$ \chg{a time derivative or an operator representing 
a convolution in time with a friction kernel}, $A$ 
a matrix of elastic constants, $f(t)$ the external force and $\eta(t)$ some zero-mean Gaussian noise. 
To demonstrate the independence with respect to the force $f(t)$, we first average the above equation
\begin{equation}
\gamma \ddt \langle x(t)\rangle = -A \langle x(t)\rangle + f(t)
\end{equation}
If we then subtract this equation from the original, introducing $y(t)=x(t)-\langle x(t)\rangle$, we get
\begin{equation}
\gamma \ddt y(t) = -A y(t) + \eta(t)
\end{equation}
Thus the probability distribution for $y$ does not depend on the external force. It will be a Gaussian of the form
\begin{equation}
P_y(y,t)= \frac{1}{\sqrt{\det(2\pi V(t))}}\exp\left(-\frac{1}{2}y^T V^{-1}(t)y\right)
\end{equation}
where $V(t)=\langle y(t)y^T(t)\rangle$ is the covariance matrix of $x(t)$. The \chg{stochastic entropy of the system 
at time $t$, $S(t)$, is
\begin{equation}
S(t)=-\ln P_y(x-\langle x(t)\rangle,t).
\end{equation}
and we deduce that the average of this, also called the Shannon entropy, is
\begin{align}
\langle S(t) \rangle &=-\langle \ln P_y(x-\langle x(t)\rangle,t)\rangle \\
&=\frac{1}{2}\ln \det(2\pi V(t))+\frac{1}{2}\langle y^T V^{-1}(t)y\rangle\\
&=\frac{1}{2}\ln \det(2\pi V(t))+\frac{1}{2}{\rm Tr}\, I
\label{Shanon}
\end{align}
where ${\rm Tr}\, I$ is the number of degrees of freedom ({\rm i.e.} components of $x(t)$).} 

Importantly, we \chg{can conclude from the above calculation that the Shannon entropy $\langle S(t)\rangle$} is independent 
of $f(t)$, since $V(t)$ does not depend on $f(t)$. \chg{Note that this conclusion holds both at the level of the 
full chain where dynamics is Markovian, as well as in the effective non-Markovian descriptions where $\ddt$ corresponds 
to a convolution with a memory kernel.
Also note that, a}s common sense would also suggest, when comparing the case of equidistant particles vs. random
equilibrium condition, given that the variance of the former is smaller than that of the latter, 
the entropy \chg{$\langle S(t)\rangle $} is smaller for the case of equidistant particles since there 
is more order or less uncertainty in the knowledge of the positions of the particles.

\section{Two particle example involving negative dissipation rate}\label{Appendix_B}

\begin{figure}
\includegraphics[width=8cm]{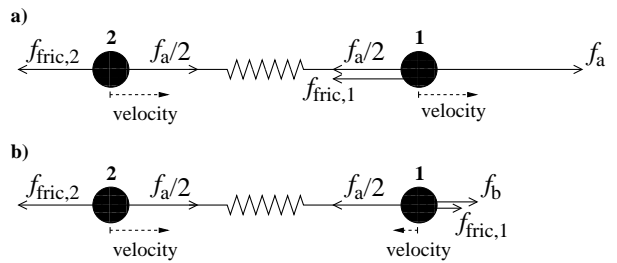}
\caption{Sketches of the two particle system at times $t=t^*+0^-$ (a) and $t=t^*+0^+$ (b) for a case with $0<f_b<f_a/2$. The velocities and frictional 
forces $f_{{\rm fric,}i}=-\xi{\dot x}_i$ are assumed to be close to their noise-averaged values.} \label{fig:two_sketch}
\end{figure}

\begin{figure}[t]
\includegraphics[width=8cm]{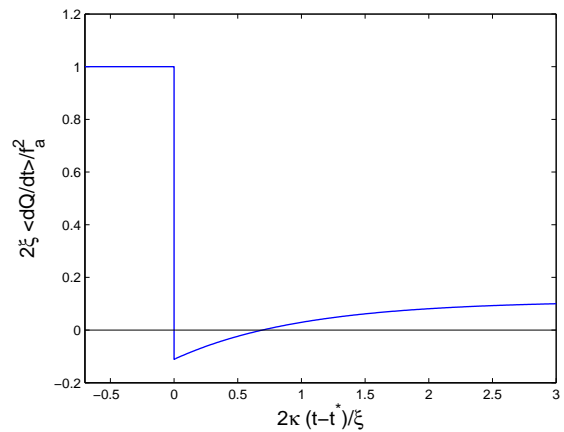}
\caption{Plot of the average dissipation rate $2\xi \langle {\dot Q} \rangle /f_a^2$ 
from Eq. (\ref{eq:Qdot_1p}) as a function of $2\kappa(t-t^*)/\xi$ for $f_b/f_a=1/3$ 
} \label{fig:Wdot_1p}
\end{figure}

As a simple illustration of the origin of negative dissipation rate, let us consider two overdamped 
particles connected by a harmonic attraction as depicted in Fig. \ref{fig:two_sketch}. 
The two particles equations of motion are a reduced version of the 
discrete system in Section \ref{sec:III}
\begin{align}
\xi\frac{d x_1}{d t}&=-\kappa\left[x_1-x_2\right]+\eta_1+f_{\rm ext}(t)\label{eq:eom1}\\
\xi\frac{d x_2}{d t}&=-\kappa\left[x_2-x_1\right]+\eta_2
\end{align}
with $L=\infty$ and an external force $f_{\rm ext}(t)$ acting only on particle 1. This system also has an 
effective description for particle 1, in which particle 2 is integrated out. 
To show this, we introduce the Laplace transform of $x_1(t)$ (resp. $x_2(t)$) as $x_1(s)$ (resp. $x_2(s)$), 
and similarly for the noises $\eta_1$, $\eta_2$ and $f_{\rm ext}$. 
By solving the equation of motion for particle 2 and inserting it in Eq. (\ref{eq:eom1}), one finds:
\be
K(s) [s x_1(s)-x_1(t=0)]= \eta_{\rm eff}(s) + f_{\rm ext}(s),
\label{eq LP}
\ee
where $K(s)=\xi + \kappa/(s + \kappa/\xi)$ and the effective noise
depends on the noises $\eta_1$, $\eta_2$ and on the initial conditions as
\be
\eta_{\rm eff}(s)= 
\eta_1(s)
+\frac{\kappa}{\xi s + \kappa} \left[ \eta_2(s) + \xi (x_2(t=0)-x_1(t=0)) \right].
\ee
By the convolution theorem, this implies the generalized Langevin equation
\begin{equation}
\int_0^t K(t-t')\frac{d x_1}{d t'}\chg{d t'}=\eta_{\rm eff}(t)+f_{\rm ext}(t)\label{eq:B3},
\end{equation}
in which the friction kernel $K(t)$, which is the inverse Laplace transform of $K(s)$, is given by
\begin{equation}
K(t)=2\xi \delta(t)+\kappa e^{-\kappa t/\xi}.
\end{equation}
The factor two arises since we define that the delta function only contributes half 
its unity mass when it is at the boundary point of an integral. Assuming that the relative distance 
$x_1-x_2$ is initially at thermal equilibrium with respect to the harmonic potential 
then the effective noise can be shown to satisfy the fluctuation-dissipation relation \cite{kubo66}
\begin{equation}
\langle \eta_{\rm eff}(t)\eta_{\rm eff}(t')\rangle = k_B T K(|t-t'|) 
\end{equation}
Notice that there is no potential in Eq. (\ref{eq:B3}) and thereby no internal energy in the effective description.
Therefore, work and heat must be equal in that description, even at the trajectory level, i.e., 
${\dot Q}={\dot W}=f_{\rm ext}{\dot x}_1$.

A simple example of a situation where the mean dissipation rate, $\langle{\dot Q}\rangle$ temporarily 
becomes negative can be found by considering an external force, which is constant except for a jump 
at a time $t^*$:
\begin{equation}
f_{\rm ext}(t)=\left\{\begin{array}{l l}f_a,&\quad t<t^*\\
f_b. &\quad t>t^*\end{array} \right.
\end{equation}
If we assume that the system is in a state with stationary velocity before time $t^*$, then the 
average velocity must make the total frictional force balance $f_a$. Thus the average velocity must be
\begin{equation}
\langle {\dot x}_1(t) \rangle =\frac{f_a}{2\xi},\quad t<t^*
\label{v short time}
\end{equation}
and the average distance between the two particles must be such that the harmonic spring force balance the friction 
on each particle which has size $f_a/2$, i.e., the distance between the particles is $\langle x_1-x_2\rangle=f_a/(2\kappa)$.
This is confirmed by an explicit calculation, using Eq.~(\ref{eq LP}) in Laplace space. For that, 
one needs the Laplace transform of the force $f_{\rm ext}(t)$, which is 
$f_{\rm ext}(s)=[f_a (1 - e^{-s t^*}) + f_b e^{-s t^*}]/s$.
After averaging over noise, and performing an
inverse Laplace transform, one indeed finds Eq.~(\ref{v short time}) when the time $t$ is  
such that $t < t^*$ but provided it is also much longer than the relaxation time of the spring 
$t \gg \xi/(2 \kappa)$.

Now, if the external force switches to a value $f_b$ with $f_b<f_a/2$ (assuming they are both positive), 
then the harmonic force will win out for a while, pulling particle 1 in the opposite direction of $f_a$. 
The distance between the particles will relax exponentially to the new stationary value $f_b/(2\kappa)$ 
with a rate constant $2\kappa/\xi$. Thus the velocity of particle 1 will be
\begin{equation}
\label{average vel}
\langle {\dot x}_1(t) \rangle=\frac{f_b}{2\xi}-\frac{f_a-f_b}{2\xi}e^{-2\kappa (t-t^*)/\xi},\quad t>t^*
\end{equation}
and we therefore obtain an average dissipation rate
\begin{equation}
\langle {\dot Q} \rangle =\left\{\begin{array}{l l}\frac{f_a^2}{2\xi},&\quad t<t^*\\
\frac{f_b}{2\xi}\left[f_b-(f_a-f_b)e^{-2\kappa (t-t^*)/\xi}\right],&\quad t>t^*\end{array}\right.\label{eq:Qdot_1p}
\end{equation}
which is temporarily negative if $0<f_b<f_a/2$. 
An example where $\langle {\dot Q} \rangle$ is temporarily negative is plotted in Fig. \ref{fig:Wdot_1p}.

\chg{We note that the negativity of $\langle {\dot Q} \rangle$ implies that $\langle {\dot S}_{\rm tot} 
\rangle$ can also become negative. To see this first note that the proof of $\Delta S_m=Q/T$ in Subsection 
\ref{subsec:sec_law} does not assume a specific form of the friction kernel in the generalized Langevin equation, 
and it therefore applies also to the present case. Therefore we have
\begin{equation}
\langle {\dot S}_{\rm tot} \rangle=\langle {\dot S} \rangle +\langle {\dot S}_{\rm m} \rangle 
= \langle {\dot S} \rangle + \langle {\dot Q} \rangle/T.
\end{equation}
Secondly, as shown in Appendix \ref{Appendix_A} the averaged Shannon entropy $\langle {S} \rangle$ 
is independent of the external force. Therefore, whatever the value of the term $\langle {\dot S} \rangle$ 
is at time $t=t^*+0^+$ then $\langle {\dot Q} \rangle/T$ can always be made more negative with a larger magnitude 
at this point in time by increasing the external forces $f_a$ and $f_b$. Thus $\langle {\dot S}_{\rm tot} \rangle$ 
can be made negative at $t=t^*+0^+$.
}

\chg{One can also check that this result is consistent with section \ref{subsec:stochent}. Since the statistics of the 
tagged particle is also Gaussian in this two particle example, 
the derivations in the subsection holds up to and including Eq.~(\ref{S_nm}). Then, using Eq.~(\ref{eq LP})
one can calculate the variance of the position of particle 1, deduce from it the time dependent diffusion coefficient 
$\Dt$ and together with the average velocity of Eq.~(\ref{average vel}), 
one can evaluate the non-Markovian contribution to the entropy production $\langle {\dot S}_{nM} \rangle$
introduced in Eq.~(\ref{S_nm}). As expected, this term can become negative.} 

\bibliographystyle{myapsrev}
\bibliography{rouse_refs}

\end{document}